\documentclass[twocolumn,superscriptaddress,letter]{revtex4}%
\usepackage{amssymb}
\usepackage{color}
\usepackage{graphicx}
\usepackage{dcolumn}
\usepackage{bm}
\usepackage[header,title,page,titletoc]{appendix}
\usepackage{amsmath}
\usepackage{amsfonts}%
\providecommand{\U}[1]{\protect \rule{.1in}{.1in}}

\begin{document}
\title{Defective Edge states and Anomalous Bulk-boundary Correspondence for
Topological Insulators under Non-Hermitian Similarity Transformation}
\author{Can Wang}
\affiliation{Center for Advanced Quantum Studies, Department of Physics, Beijing Normal
University, Beijing 100875, China}
\author{Xiao-Ran Wang}
\affiliation{Center for Advanced Quantum Studies, Department of Physics, Beijing Normal
University, Beijing 100875, China}
\author{Cui-Xian Guo}
\affiliation{Center for Advanced Quantum Studies, Department of Physics, Beijing Normal
University, Beijing 100875, China}
\author{Su-Peng Kou}
\thanks{Corresponding author}
\email{spkou@bnu.edu.cn}
\affiliation{Center for Advanced Quantum Studies, Department of Physics, Beijing Normal
University, Beijing 100875, China}

\begin{abstract}
It was known that for non-Hermitian topological systems due to the
non-Hermitian skin effect, the bulk-edge correspondence is broken down. In
this paper, by using one-dimensional Su-Schrieffer-Heeger model and
two-dimensional (deformed) Qi-Wu-Zhang model as examples, we focus on a
special type of non-Hermitian topological system without non-Hermitian skin
effect -- topological systems under non-Hermitian similarity transformation.
In these non-Hermitian systems, the defective edge states and the breakdown of
bulk-edge correspondence are discovered. To characterize the topological
properties, we introduce a new type of inversion symmetry-protected
topological invariant -- total $Z_{2}$ topological invariant. In topological
phases, defective edge states appear. With the help of the effective edge
Hamiltonian, we find that the defective edge states are protected by
(generalized) chiral symmetry and thus the (singular) defective edge states
are unstable against the perturbation breaking the chiral symmetry. In
addition, the results are generalized to non-Hermitian topological insulators
with inversion symmetry in higher dimensions. This work could help people to
understand the defective edge states and the breakdown of bulk-edge
correspondence for non-Hermitian topological systems.

\end{abstract}

\maketitle

\section{Introduction}

Non-Hermitian topological
systems\cite{Rudner2009,Esaki2011,Hu2011,Liang2013,Zhu2014,Lee2016,San2016,Leykam2017,Shen2018,Lieu2018,
Xiong2018,Kawabata2018,Gong2018,Yao2018,YaoWang2018,Yin2018,Kunst2018,KawabataUeda2018,Alvarez2018,
Jiang2018,Ghatak2019,Avila2019,Jin2019,Lee2019,Liu2019,38-1,38,chen-class2019,Edvardsson2019,
Herviou2019,Yokomizo2019,zhouBin2019,Kunst2019,Deng2019,SongWang2019,xi2019,Longhi2019,chen-edge2019,Wang2020,Sato2012,Schomerus2013,Malzard2015,Harter2016,Xu2017,Menke2017,Takata2018,KawabataU2019,Okugawa2019,Budich2019,Yang2019,Lin2019,ZhangK2019,WangH2019,Carlstrom2019,LonghiS2019,LeeC2019,RudnerM,ZengQ,WuH2019,LinS2019}
have been confirmed with diverse peculiar underlying physics distinguishing
from its Hermitian counterparts\cite{Kane2010,Qi2011,Ali2012,chi2016,ban2016}.
Generally, the non-Hermitian terms i.e., imaginary-mass, or
imaginary-momentum, or anti-commutating in tight-binding topological model
bring novel properties such as complex spectra, defective edge
states\cite{Lee2016,Yin2018}, non-Hermitian skin
effect\cite{Yao2018,Ghatak2019,Lee2019,SongWang2019,Longhi2019}, ... Due to
the conventional bulk-boundary correspondence (BBC) collapsed in non-Hermitian
topological
insulators\cite{Xiong2018,Yao2018,YaoWang2018,Kunst2018,Herviou2019,Yokomizo2019,Kunst2019,Deng2019,Longhi2019}%
, the non-Bloch topological invariants\cite{Yao2018} and the effective theory
for the edge states\cite{Wang2020} were introduced to describe the
non-Hermitian topological systems. Within the framework of Altland-Zirnbauer
(AZ) theory, the topological invariants classifications based on different
symmetries in the non-Hermitian topological systems were
developed\cite{Gong2018,38,38-1}. After considering reflection symmetry, the
classification of non-Hermitian topological systems is also
finished\cite{chen-class2019}.

However, a large class of non-Hermitian topological systems with real energy
spectra have not been discussed:\emph{ topological system under non-Hermitian
similarity transformation.} In particular, under non-Hermitian similarity
transformation, the energy levels are all real and same to those of the
Hermitian counterparts\cite{Fernandez2016,Rui2019}. So, there doesn't exist
the non-Hermitian skin effect. However, the quantum states of these systems
may be quite different from that of the Hermitian counterpart. Here, to
completely understand the underlying physics of this type of non-Hermitian
topological systems, we ask the following questions:

\begin{enumerate}
\item What are the\emph{ topological invariants} in the bulk for this type of
non-Hermitian topological systems?

\item Does there exist the defective edge states in this type of non-Hermitian
topological systems \emph{without} non-Hermitian skin effect?

\item How to \emph{accurately} characterize the physics of edge states and the
possible anomalous BBC of defective edge states?

\item Are the defective edge states in this non-Hermitian topological system
\emph{stable}?
\end{enumerate}

In this paper, to answer above questions, we systematically study the
non-Hermitian topological systems by investigating one-dimensional (1D)
Su-Schrieffer-Heeger (SSH) model and two-dimensional (2D) Qi-Wu-Zhang (QWZ)
model under non-Hermitian similarity transformation. To characterize these
non-Hermitian topological systems, we introduce a new type of
symmetry-protected topological invariant -- total $Z_{2}$ topological
invariant. In topological phases, defective edge states appear, i.e., edge
states on the ends of finite non-Hermitian topological systems with
non-Hermitian coalescence. To accurately characterize the physics of defective
edge states and the anomalous BBC, the effective edge Hamiltonians are
obtained. With the help of the effective edge Hamiltonian, we find that the
defective edge states are protected by (generalized) chiral symmetry and the
(singular) defective edge states are unstable against the perturbation
breaking the chiral symmetry. In addition, we generalize the results to
non-Hermitian topological insulators in higher dimensions.

This paper is organized as follows. In Sec. II, we explore the defective edge
states and the anomalous BBC for 1D Su-Schrieffer-Heeger under non-Hermitian
similarity transformation. In Sec. III, we study the defective edge states and
the anomalous BBC for 2D Chern insulator under non-Hermitian similarity
transformation. In Sec. IV we generalize the theory to topological insulators
in higher dimensions under non-Hermitian similarity transformation. In the
end, we give a brief conclusion in Sec. V.

\section{Anomalous BBC and defective edge states for 1D Su-Schrieffer-Heeger
under non-Hermitian similarity transformation}

\subsection{1D Hermitian Su-Schrieffer-Heeger model}

We begin with the simplest 1D Topological insulator -- Hermitian SSH
model,\ of which tight-binding Hamiltonian for the finite system with $N$
pairs of lattice sites is given by
\begin{equation}
H_{0}^{\mathrm{SSH}}=t_{1}\sum_{n=1}^{N}\left \vert n,B\right \rangle
\left \langle n,A\right \vert +t_{2}\sum_{n=1}^{N-1}\left \vert
n+1,A\right \rangle \left \langle n,B\right \vert +h.c.
\end{equation}
where $A~(B)$ represents the sublattices and $n$ indicates the $n$-th cell of
the lattice. Correspondingly, under periodic boundary condition, its Bloch
Hamiltonian becomes
\begin{align}
H_{0}^{\mathrm{SSH}}(k)  &  =\left(  t_{1}+t_{2}\cos k\right)  \sigma
_{x}+\left(  t_{2}\sin k\right)  \sigma_{y}\\
&  =\left(
\begin{array}
[c]{cc}%
0 & t_{1}+t_{2}e^{-ik}\\
t_{1}+t_{2}e^{ik} & 0
\end{array}
\right)  ,\nonumber
\end{align}
where the real parameters $t_{1},$ $t_{2}$ are the intra (inter)-hopping
amplitude, $\sigma_{i}$ refers to the Pauli matrices and the $2\times2$ matrix
above is defined as the bulk Hamiltonian $H_{0}^{\mathrm{SSH}}(k).$ Under
periodic boundary condition, the energy spectra are
\begin{equation}
E(k)=\pm \sqrt{\left(  t_{1}+t_{2}\cos k\right)  ^{2}+\left(  t_{2}\sin
k\right)  ^{2}}.
\end{equation}
Fig. 1(b) shows the energy spectra of SSH model.

\subsubsection{Chiral symmetry and inversion symmetry}

The Hamiltonian $H_{0}^{\mathrm{SSH}}(k)$ possesses chiral symmetry, i.e.
\begin{equation}
\sigma_{z}H_{0}^{\mathrm{SSH}}\left(  k\right)  \sigma_{z}=-H_{0}%
^{\mathrm{SSH}}\left(  k\right)  .
\end{equation}
$H_{0}^{\mathrm{SSH}}\left(  k\right)  $ also has inversion symmetry
satisfying the relation
\begin{equation}
\mathcal{\hat{I}}H_{0}^{\mathrm{SSH}}\left(  k\right)  \mathcal{\hat{I}}%
^{-1}=H_{0}^{\mathrm{SSH}}\left(  -k\right)  ,
\end{equation}
where the inversion operator $\mathcal{\hat{I}}$ is $\sigma_{x}$. This
equation implies
\begin{equation}
E(k)=E(-k).
\end{equation}

\subsubsection{Total $Z_{2}$ topological invariant}

According to AZ classification, it is 1D AIII type, of which the topological
invariant is a winding number%
\begin{equation}
w=\frac{1}{2\pi}\int_{-\pi}^{\pi}\partial_{k}\phi_{n}(k)\cdot dk.
\end{equation}
where $\phi_{n}(k)=\tan^{-1}(d_{y}/d_{x})$ with $d_{y}=t_{2}\sin k$ and
$d_{x}=t_{1}+t_{2}\cos k$. There are two phases, topological phase with $w=1$
in the region of $\left \vert t_{1}\right \vert <\left \vert t_{2}\right \vert $
and trivial phase with $w=0$ in the region of $\left \vert t_{1}\right \vert
>\left \vert t_{2}\right \vert $. At $\left \vert t_{1}\right \vert =\left \vert
t_{2}\right \vert ,$ there exists a topological phase transition. See the phase
diagram in Fig. 1(a) ($\beta=0$ case).

However, in this paper, to characterize the SSH model with chiral symmetry and
inversion symmetry, instead of winding number $w$, we introduce a new
topological invariant -- total $Z_{2}$ topological invariant $\eta
=\eta_{k\mathbf{=}0}\eta_{k\mathbf{=}\pi}$. See the detailed definition in
below discussion.

Now, the Bloch Hamiltonian is divided into three parts
\begin{align}
H_{0}^{\mathrm{SSH}}(k)  &  =H_{0}^{\mathrm{SSH}}(k\neq0/\pi)+H_{0}%
^{\mathrm{SSH}}(k=0)\nonumber \\
+H_{0}^{\mathrm{SSH}}(k  &  =\pi)
\end{align}
Here, $k=0/\pi$ are the high symmetry points in momentum space. The quantum
states at these high symmetry points are invariant under inversion operation,
i.e.,
\begin{align}
\mathcal{\hat{I}}\left \vert \psi(k=0)\right \rangle  &  =\left \vert
\psi(k=0)\right \rangle ,\\
\mathcal{\hat{I}}\left \vert \psi(k=\pi)\right \rangle  &  =\left \vert
\psi(k=\pi)\right \rangle .\nonumber
\end{align}
For $k=0$, we have
\begin{equation}
\frac{1}{2}\mathrm{Tr}[\mathcal{\hat{I}}\cdot H_{0}^{\mathrm{SSH}}%
(k=0)]=t_{1}+t_{2};
\end{equation}
for $k=\pi$, we have
\begin{equation}
\frac{1}{2}\mathrm{Tr}[\mathcal{\hat{I}}\cdot H_{0}^{\mathrm{SSH}}%
(k=\pi)]=t_{1}-t_{2}.
\end{equation}

To describe the topological structure of $H_{0}^{\mathrm{SSH}}(k)$, we define
two $Z_{2}$ topological invariants,
\begin{equation}
\eta_{k\mathbf{=}0}=\frac{\mathrm{Tr}[\mathcal{\hat{I}}\cdot H_{0}%
^{\mathrm{SSH}}[k=0\mathbf{]]}}{\left \vert \mathrm{Tr}[\mathcal{\hat{I}}\cdot
H_{0}^{\mathrm{SSH}}[k=0\mathbf{]]}\right \vert }=\frac{t_{1}+t_{2}}{\left \vert
t_{1}+t_{2}\right \vert },
\end{equation}
and
\begin{equation}
\eta_{k\mathbf{=}\pi}=\frac{\mathrm{Tr}[\mathcal{\hat{I}}\cdot H_{0}%
^{\mathrm{SSH}}[k=\pi \mathbf{]]}}{\left \vert \mathrm{Tr}[\mathcal{\hat{I}%
}\cdot H_{0}^{\mathrm{SSH}}[k=\pi \mathbf{]]}\right \vert }=\frac{t_{1}-t_{2}%
}{\left \vert t_{1}-t_{2}\right \vert }%
\end{equation}
where $\mathcal{\hat{I}}=\sigma_{x}$. Now, we use the number $^{\prime
}1^{\prime}$ to denote the case $\eta_{k\mathbf{=}0}=-1$ or $\eta
_{k\mathbf{=}\pi}=-1$\ and the number $^{\prime}0^{\prime}$ to denote the case
$\eta_{k\mathbf{=}0}=1$ or $\eta_{k\mathbf{=}\pi}=1.$ Hence, there are $4$
different universal classes of topological phases denoted by $\left(
11\right)  $,\ $\left(  00\right)  $,\ $\left(  10\right)  $,\ $\left(
01\right)  $\cite{Liu2011}. The following table shows the $Z_{2}$ topological
invariants of different universal classes of topological phases%
\[%
\begin{array}
[c]{lllll}
& \left(  11\right)  & \left(  10\right)  & \left(  01\right)  & \left(
00\right) \\
k=0 & -1 & -1 & 1 & 1\\
k=\pi & -1 & 1 & -1 & 1
\end{array}
\]

The total $Z_{2}$ topological invariant is defined as
\begin{equation}
\eta=\eta_{k\mathbf{=}0}\eta_{k\mathbf{=}\pi}=\left \{
\begin{array}
[c]{l}%
+1,\text{ trivial phase}\\
-1,\text{ topological phase}%
\end{array}
\right.  .
\end{equation}
$\eta$ becomes a topological invariant to characterize the universal
properties of different topological phases for SSH model with inversion
symmetry. There are $2$ trivial phases: $\left(  11\right)  $, $\left(
00\right)  $,\textrm{ }and $2$ topological phases: $\left(  10\right)
$,\ $\left(  01\right)  $. In particular, the phase diagram from the $Z_{2}$
topological invariant $\eta$ and that from the winding number $\omega$ are
same, i.e., topological phases in the region of $\left \vert t_{1}\right \vert
<\left \vert t_{2}\right \vert $ and trivial phases in the region of $\left \vert
t_{1}\right \vert >\left \vert t_{2}\right \vert $.

\begin{figure}[ptb]
\includegraphics[clip,width=0.48\textwidth]{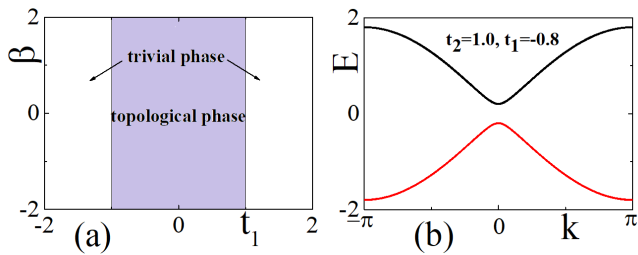}\caption{(Color online)
(a) Global phase diagram for 1D SSH model under non-Hermitian similarity
transformation. The purple area represents the topological phase with
$\eta=-1$, while the white space represents the trivial phase with $\eta=1$,
respectively. The topological phase transition occurs at $t_{1}=t_{2}$£»
(b) The energy spectra for 1D SSH model under non-Hermitian similarity
transformation. The energy spectra are independent on $\beta.$}%
\end{figure}

\subsubsection{Effective edge Hamiltonian}

The non-zero winding number or total $Z_{2}$ topological invariant $\eta$
guarantees the edge states with zero energy for a system in thermodynamical
limit, $N\rightarrow \infty$. For the case of non-topological phase, the edge
states disappear. This leads to the conventional bulk-boundary correspondence
for Hermitian topological systems.

In topological phase, we consider the edge states $\left \vert \mathrm{e}%
_{0}^{\text{L}}\right \rangle $ and $\left \vert \mathrm{e}_{0}^{\text{R}%
}\right \rangle $ on the left and right ends of the semi-infinite chain as the
basis\cite{si}, i.e.,%
\begin{equation}
\left \vert \mathrm{e}_{0}^{\text{L}}\right \rangle =\frac{1}{\mathcal{N}}%
\sum_{n=1}^{N-1}(\frac{t_{1}}{t_{2}})^{n-1}|n\rangle \otimes(1,0)^{\text{T}}%
\end{equation}
and
\begin{equation}
|\mathrm{e}_{0}^{\text{R}}\rangle=\frac{1}{\mathcal{N}}\sum_{n=0}^{N-1}%
(\frac{t_{1}}{t_{2}})^{n}|N-n\rangle \otimes(0,1)^{\text{T }},
\end{equation}
where the normalization factor is
\begin{equation}
\mathcal{N=}\sqrt{\left(  1-\left(  \frac{t_{1}}{t_{2}}\right)  ^{2N}\right)
/\left(  1-\left(  \frac{t_{1}}{t_{2}}\right)  ^{2}\right)  }.
\end{equation}
$(1,0)^{\text{T}}$ and $(0,1)^{\text{T}}$ denote the state vectors of
two-sublattices. For an semi-infinite chain, we have\ $N\rightarrow \infty$. In
general, the wave-function for an edge state can be written as a superposition
of the two end states at the left and right ends
\begin{equation}
\left \vert \mathrm{\psi}_{0}\right \rangle =C_{1}\left \vert \mathrm{e}%
_{0}^{\text{L}}\right \rangle +C_{2}\left \vert \mathrm{e}_{0}^{\text{R}%
}\right \rangle
\end{equation}
where $C_{1}$\textrm{ }and $C_{2}$ are complex numbers, and $\left \vert
C_{1}\right \vert ^{2}+\left \vert C_{2}\right \vert ^{2}=1$.

To characterize the two edge states, we introduce an effective edge
Hamiltonian,
\begin{equation}
\mathcal{\hat{H}}_{\mathrm{eff}}=\left(
\begin{array}
[c]{cc}%
\varepsilon_{LL} & \varepsilon_{LR}\\
\varepsilon_{RL} & \varepsilon_{RR}%
\end{array}
\right)
\end{equation}
where%
\begin{align}
\varepsilon_{LL}  &  =\left \langle \mathrm{e}_{0}^{\text{L}}\right \vert
H_{0}^{\mathrm{SSH}}\left \vert \mathrm{e}_{0}^{\text{L}}\right \rangle ,\text{
}\varepsilon_{LR}=\left \langle \mathrm{e}_{0}^{\text{L}}\right \vert
H_{0}^{\mathrm{SSH}}\left \vert \mathrm{e}_{0}^{\text{R}}\right \rangle ,\\
\varepsilon_{RL}  &  =\left \langle \mathrm{e}_{0}^{\text{R}}\right \vert
H_{0}^{\mathrm{SSH}}\left \vert \mathrm{e}_{0}^{\text{L}}\right \rangle ,\text{
}\varepsilon_{RR}=\left \langle \mathrm{e}_{0}^{\text{R}}\right \vert
H_{0}^{\mathrm{SSH}}\left \vert \mathrm{e}_{0}^{\text{R}}\right \rangle
.\nonumber
\end{align}
For this model, we have
\begin{align}
\varepsilon_{LL}  &  =-\varepsilon_{RR}=0,\\
\varepsilon_{LR}  &  =\varepsilon_{RL}=\Delta=\frac{\left(  t_{2}^{2}%
-t_{1}^{2}\right)  }{t_{2}}(\frac{t_{1}}{t_{2}})^{N}.\nonumber
\end{align}
The effective edge Hamiltonian becomes%
\begin{equation}
\mathcal{\hat{H}}_{\mathrm{eff}}=\Delta \cdot \tau^{x}%
\end{equation}
where $\tau^{i}$ is the Pauli matrix acting on the two edge states. As a
result, the energy levels for the two edge states are%
\begin{equation}
E_{\pm}=\pm \frac{\left(  t_{2}^{2}-t_{1}^{2}\right)  }{t_{2}}(\frac{t_{1}%
}{t_{2}})^{N}%
\end{equation}
and their eigenstates are
\begin{equation}
\frac{1}{\sqrt{2}}(\left \vert \mathrm{e}_{0}^{\text{L}}\right \rangle
+\left \vert \mathrm{e}_{0}^{\text{R}}\right \rangle ),\text{ }\frac{1}{\sqrt
{2}}(\left \vert \mathrm{e}_{0}^{\text{L}}\right \rangle -\left \vert
\mathrm{e}_{0}^{\text{R}}\right \rangle ).
\end{equation}

\subsection{1D SSH model under non-Hermitian similarity transformation}

In the second step, we consider the SSH model under the non-Hermitian
similarity transformation,
\begin{equation}
S=\left(
\begin{array}
[c]{cc}%
{1} & 0\\
0 & e^{\beta}%
\end{array}
\right)
\end{equation}
where $\beta$ denotes the non-Hermiticity strength. Under the non-Hermitian
similarity transformation, we have%
\begin{align}
\sigma_{x}  &  =\left(
\begin{array}
[c]{cc}%
{0} & 1\\
1 & {0}%
\end{array}
\right)  \rightarrow(\sigma_{x})^{\beta}=S^{-1}\sigma_{x}S\\
&  =\cosh(\beta)\sigma_{x}+i\sinh(\beta)\sigma_{y}=\left(
\begin{array}
[c]{cc}%
{0} & e{^{\beta}}\\
e^{-\beta} & {0}%
\end{array}
\right)  ,\nonumber
\end{align}

\begin{align}
\sigma_{y}  &  =\left(
\begin{array}
[c]{cc}%
{0} & -i\\
i & {0}%
\end{array}
\right)  \rightarrow(\sigma_{y})^{\beta}=S^{-1}\sigma_{y}S\\
&  =\cosh(\beta)\sigma_{y}-i\sinh(\beta)\sigma_{x}=\left(
\begin{array}
[c]{cc}%
{0} & -ie{^{\beta}}\\
ie^{-\beta} & {0}%
\end{array}
\right)  ,\nonumber
\end{align}
and
\begin{equation}
\sigma_{z}\rightarrow(\sigma_{z})^{\beta}=S^{-1}\sigma_{z}S=\sigma_{z}.
\end{equation}

As a result, under the non-Hermitian similarity transformation, the Bloch
Hamiltonian of SSH model turns into a nonreciprocal one, that is
\begin{align}
H_{0}^{\mathrm{SSH}}  &  \rightarrow H^{\mathrm{SSH}}=S^{-1}H_{0}%
^{\mathrm{SSH}}S\\
&  =\left(
\begin{array}
[c]{cc}%
0 & e{^{\beta}}(t_{1}+t_{2}e^{-ik})\\
e{^{-\beta}}(t_{1}+t_{2}e^{ik}) & 0
\end{array}
\right) \nonumber \\
&  =\left(  t_{1}\cosh \beta+t_{2}\cos(k+i\beta)\right)  \sigma_{x}\nonumber \\
&  +\left(  i\text{ }t_{1}\sinh \beta+t_{2}\sin(k+i\beta)\right)  \sigma
_{y}.\nonumber
\end{align}
To describe the edge states, we rewrite the Hamiltonian in real space
\begin{align}
H^{\mathrm{SSH}}  &  =\sum_{n}t_{{1}L}|n,A\rangle \langle n,B|+t_{{2}%
R}|n,A\rangle \langle n-1,B|\nonumber \\
&  +t_{{1}R}|n,B\rangle \langle n,A|+t_{{2}L}|n,B\rangle \langle n+1,A|,
\end{align}
where $t_{{1(2)L(R)}}$ is the right (left) intra (inter)-hopping amplitude.
Then, the effective hopping parameters in real space become
\begin{align}
t_{1L}  &  =t_{1}e{^{\beta},}\text{ }t_{1R}=t_{1}e{^{-\beta},}\\
t_{2L}  &  =t_{2}e{^{-\beta},}\text{ }t_{2R}=t_{2}e{^{\beta}}.\nonumber
\end{align}

The eigenvalues of the Hamiltonian $H^{\mathrm{SSH}}$ for the nonreciprocal
SSH model are same to those of the Hermitian SSH model, i.e.,
\begin{equation}
E_{\pm}\left(  k\right)  =\pm \sqrt{\left(  t_{1}+t_{2}\cos k\right)
^{2}+\left(  t_{2}\sin k\right)  ^{2}}.
\end{equation}
Fig. 1(b) also shows the energy spectra of SSH model under non-Hermitian
similarity transformation that are all real and unchanged with $\beta$. As a
result, due to the real energy spectra, there doesn't exist the non-Hermitian
skin effect.

\subsection{Total $Z_{2}$ topological invariant}

For SSH model under non-Hermitian similarity transformation, the Hamiltonian
also has chiral symmetry, i.e., $\sigma_{z}H^{\mathrm{SSH}}\left(  k\right)
\sigma_{z}=-H^{\mathrm{SSH}}\left(  k\right)  $ and inversion symmetry, i.e.,
$\mathcal{\hat{I}}H^{\mathrm{SSH}}\left(  k\right)  \mathcal{\hat{I}}%
^{-1}=H^{\mathrm{SSH}}\left(  -k\right)  $ where the inversion operator
$\mathcal{\hat{I}}$ is $(\sigma_{x})^{\beta}$.

For the non-Hermitian topological system with inversion symmetry, we use the
total $Z_{2}$ topological invariant $\eta$ to characterize its topological
properties
\begin{align}
\eta &  =\eta_{k\mathbf{=}0}\eta_{k\mathbf{=}\pi}\nonumber \\
&  =\frac{\mathrm{Tr}[[\mathcal{\hat{I}}\cdot H_{0}^{\mathrm{SSH}%
}[k=0\mathbf{]]}}{\left \vert \mathrm{Tr}[\mathcal{\hat{I}}\cdot H_{0}%
^{\mathrm{SSH}}[k=0\mathbf{)]]}\right \vert }\frac{\mathrm{Tr}[\mathcal{\hat
{I}}\cdot H_{0}^{\mathrm{SSH}}[k=\pi \mathbf{]]}}{\left \vert \mathrm{Tr}%
[\mathcal{\hat{I}}\cdot H_{0}^{\mathrm{SSH}}[k=\pi \mathbf{)]]}\right \vert
}\nonumber \\
&  =\frac{t_{1}^{2}-t_{2}^{2}}{\left \vert t_{1}^{2}-t_{2}^{2}\right \vert }.
\end{align}
There are also two phases, topological phase with $\eta=-1$ in the region of
$\left \vert t_{1}\right \vert <\left \vert t_{2}\right \vert $ and trivial phase
with $\eta=1$ in the region of $\left \vert t_{1}\right \vert >\left \vert
t_{2}\right \vert $. At $\left \vert t_{1}\right \vert =\left \vert t_{2}%
\right \vert ,$ there exists a topological phase transition. See the phase
diagram in Fig. 1(a).

\subsection{Effective edge Hamiltonian for defective edge states}

To characterize the two edge states for SSH model under non-Hermitian
similarity transformation, we calculate the effective edge Hamiltonian,
\begin{equation}
\mathcal{\breve{H}}_{\mathrm{eff}}=\left(
\begin{array}
[c]{cc}%
\varepsilon_{LL}^{\prime} & \varepsilon_{LR}^{\prime}\\
\varepsilon_{RL}^{\prime} & \varepsilon_{RR}^{\prime}%
\end{array}
\right)  ,
\end{equation}
where $\varepsilon_{LL}^{\prime}=\left \langle \mathrm{e}_{0}^{\text{L}%
}\right \vert H^{\mathrm{SSH}}\left \vert \mathrm{e}_{0}^{\text{L}}\right \rangle
,$ $\varepsilon_{LR}^{\prime}=\left \langle \mathrm{e}_{0}^{\text{L}%
}\right \vert H^{\mathrm{SSH}}\left \vert \mathrm{e}_{0}^{\text{R}}\right \rangle
,$ $\varepsilon_{RL}^{\prime}=\left \langle \mathrm{e}_{0}^{\text{R}%
}\right \vert H^{\mathrm{SSH}}\left \vert \mathrm{e}_{0}^{\text{L}}\right \rangle
,$ $\varepsilon_{RR}^{\prime}=\left \langle \mathrm{e}_{0}^{\text{R}%
}\right \vert H^{\mathrm{SSH}}\left \vert \mathrm{e}_{0}^{\text{R}}\right \rangle
.$ Here, the basis of the wave-functions of end states are same to those for
the Hermitian case\cite{chen-edge2019}, i.e., $\left \vert \mathrm{e}%
_{0}^{\text{L}}\right \rangle $ and $|\mathrm{e}_{0}^{\text{R}}\rangle.\ $For
this model, we have
\begin{align}
\varepsilon_{LL}^{\prime}  &  =-\varepsilon_{RR}^{\prime}=0,\text{ }\\
\varepsilon_{LR}^{\prime}  &  =e{^{\beta}}\Delta,\text{ }\varepsilon
_{RL}^{\prime}=e{^{-\beta}}\Delta.\nonumber
\end{align}
where
\begin{equation}
\Delta=\frac{\left(  t_{2}^{2}-t_{1}^{2}\right)  }{t_{2}}(\frac{t_{1}}{t_{2}%
})^{N}.
\end{equation}

The effective edge Hamiltonian is obtained as
\begin{equation}
\mathcal{\breve{H}}_{\mathrm{eff}}=\Delta \cdot(\tau^{x})^{\beta}%
\end{equation}
where
\begin{equation}
(\tau^{x})^{\beta}=\cosh(\beta)\tau^{x}+i\sinh(\beta)\tau^{y}=\left(
\begin{array}
[c]{cc}%
{0} & {\mathrm{e}^{\beta}}\\
{\mathrm{e}^{-\beta}} & {0}%
\end{array}
\right)  .
\end{equation}
Although, the effective edge Hamiltonian $\mathcal{\breve{H}}_{\mathrm{eff}}$
is non-Hermitian, i.e., $\mathcal{\breve{H}}_{\mathrm{eff}}\neq
(\mathcal{\breve{H}}_{\mathrm{eff}})^{\dagger}$, the energy splitting for the
edge states doesn't change.

However, the basis of the edge states changes under the non-Hermitian
similarity transformation, i.e.,
\begin{align}
\left(
\begin{array}
[c]{c}%
\left \vert \mathrm{e}_{0}^{\text{L}}\right \rangle \\
\left \vert \mathrm{e}_{0}^{\text{R}}\right \rangle
\end{array}
\right)   &  \rightarrow \left(
\begin{array}
[c]{c}%
\left \vert \mathrm{\bar{e}}_{0}^{\text{L}}\right \rangle \\
\left \vert \mathrm{\bar{e}}_{0}^{\text{R}}\right \rangle
\end{array}
\right)  =S^{-1}\left(
\begin{array}
[c]{c}%
\left \vert \mathrm{e}_{0}^{\text{L}}\right \rangle \\
\left \vert \mathrm{e}_{0}^{\text{R}}\right \rangle
\end{array}
\right) \\
&  =\left(
\begin{array}
[c]{c}%
\left \vert \mathrm{e}_{0}^{\text{L}}\right \rangle \\
{\mathrm{e}^{-\beta}}\left \vert \mathrm{e}_{0}^{\text{R}}\right \rangle
\end{array}
\right) \nonumber
\end{align}
where $\left(
\begin{array}
[c]{c}%
\left \vert \mathrm{\bar{e}}_{0}^{\text{L}}\right \rangle \\
\left \vert \mathrm{\bar{e}}_{0}^{\text{R}}\right \rangle
\end{array}
\right)  $ is the basis for the non-Hermitian case with $\beta \neq0$. As a
result,\ the eigenstates for the two edge states turn into
\begin{equation}
\left \vert \psi_{1}\right \rangle =\frac{1}{\sqrt{1+e^{-2\beta}}}(\left \vert
\mathrm{e}_{0}^{\text{L}}\right \rangle +e^{-\beta}\left \vert \mathrm{e}%
_{0}^{\text{R}}\right \rangle )
\end{equation}
and
\[
\left \vert \psi_{2}\right \rangle =\frac{1}{\sqrt{1+e^{-2\beta}}}(\left \vert
\mathrm{e}_{0}^{\text{L}}\right \rangle -e^{-\beta}\left \vert \mathrm{e}%
_{0}^{\text{R}}\right \rangle ).
\]

To characterize the non-Hermitian properties under similarity transformation,
the states overlap $\gamma$ between the two edge states is defined as
\begin{equation}
\gamma=\left \langle \psi_{2}|\psi_{1}\right \rangle =\tanh \beta.
\end{equation}
For the case of $\beta \rightarrow0$, we have
\begin{equation}
\left(
\begin{array}
[c]{c}%
\left \vert \mathrm{\bar{e}}_{0}^{\text{L}}\right \rangle \\
\left \vert \mathrm{\bar{e}}_{0}^{\text{R}}\right \rangle
\end{array}
\right)  \rightarrow \left(
\begin{array}
[c]{c}%
\left \vert \mathrm{e}_{0}^{\text{L}}\right \rangle \\
\left \vert \mathrm{e}_{0}^{\text{R}}\right \rangle
\end{array}
\right)  .
\end{equation}
Now, we have $\gamma \rightarrow0$; On the other hand, for the case of
$\beta \rightarrow \infty$, we have
\begin{equation}
\left(
\begin{array}
[c]{c}%
\left \vert \mathrm{\bar{e}}_{0}^{\text{L}}\right \rangle \\
\left \vert \mathrm{\bar{e}}_{0}^{\text{R}}\right \rangle
\end{array}
\right)  \rightarrow \left(
\begin{array}
[c]{c}%
\left \vert \mathrm{e}_{0}^{\text{L}}\right \rangle \\
0
\end{array}
\right)  .
\end{equation}
Now, we have $\gamma \rightarrow1.$ The edge states become defective: only edge
state at left or right end exists.

\begin{figure}[ptb]
\includegraphics[clip,width=0.48\textwidth]{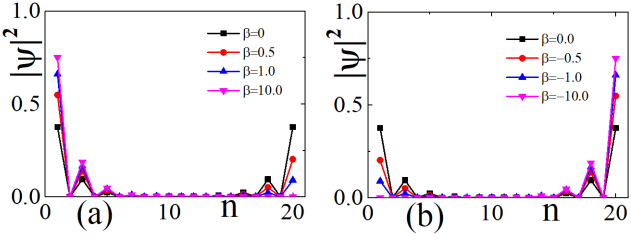}\caption{(Color online)
The wave-functions via $\beta$ for edge states of 1D SSH under non-Hermitian
similarity transformation: $\beta>0$ for (a) and $\beta<0$\ for (b). For this
case we have $t_{1}=0.5,$ $t_{2}=1.0$. }%
\end{figure}

Fig. 2(a) and Fig. 2(b) show the edge states of the SSH model under
non-Hermitian similarity transformation for case of $\beta>0$ and $\beta<0$,
respectively. In fact, the non-Hermitian similarity transformation only
polarizes the states onto one selected sublattice in the unit cell. In the
strong non-Hermitian limit $\beta \rightarrow \pm \infty,$ the edge states
localize on the A or B-sublattices that corresponds to a defective edge state
on left or right end. Fig. 3 shows the changing of bulk states via $\beta.$
From Fig. 3, one can see that there only exists the effect from sublattice
polarization, but no non-Hermitian skin effect. The weight of the bulk states
tends to accumulate on A-sublattices with increasing $\beta,$ and
simultaneously decrease on B-sublattice. Fig. 4 shows the states overlap
$\gamma$ between the two edge states$.$ From Fig. 4, one can see that the
theoretical prediction is consistent to the numerical results. When
$\beta \rightarrow0,$ the system reduces to Hermitian SSH model, the states
overlap $\gamma$ is zero, while in the strong non-Hermitian limit
$\beta \rightarrow \pm \infty,$ the states overlap $\gamma$\ turns to $1$. This
indicates the (singular) defective edge states.

\begin{figure}[ptb]
\includegraphics[clip,width=0.40\textwidth]{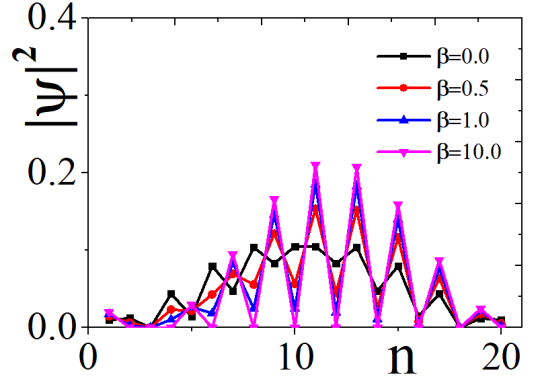}\caption{(Color online)
The wave-functions via $\beta$ for bulk states of 1D SSH under non-Hermitian
similarity transformation. For this case, we have $t_{1}=0.5,$ $t_{2}=1.0$. }%
\end{figure}

\begin{figure}[ptb]
\includegraphics[clip,width=0.40\textwidth]{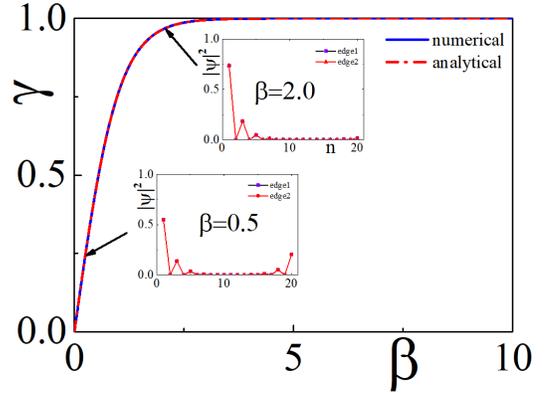}\caption{(Color online)
The state overlap $\gamma$ between the two edge states for 1D SSH under
non-Hermitian similarity transformation.}%
\end{figure}

It had been believed that the defective edge states come from the
non-Hermitian skin effect. However, for the SSH model under non-Hermitian
similarity transformation, without non-Hermitian skin effect in the bulk, the
defective edge states still exist. Therefore, this result is new.

In addition, we point out that the singular defective edge states with
$\gamma=1$ are protected by the chiral symmetry%
\begin{equation}
\sigma_{z}H^{\mathrm{SSH}}(k=0)\sigma_{z}=-H^{\mathrm{SSH}}(k=0).
\end{equation}
Due to the chiral symmetry, the wave-function of edge state on left end
localizes on A-sublattice and the wave-function of edge state on right end
localizes on B-sublattice. As a result, the non-Hermitian similarity
transformation on different sublattices, i.e., $\sigma_{x}\rightarrow
(\sigma_{x})^{\beta}=\left(
\begin{array}
[c]{cc}%
{0} & e{^{\beta}}\\
e^{-\beta} & {0}%
\end{array}
\right)  $ leads to the non-Hermitian similarity transformation on different
edge states, i.e., $\tau_{x}\rightarrow(\tau_{x})^{\beta}=\left(
\begin{array}
[c]{cc}%
{0} & e{^{\beta}}\\
e^{-\beta} & {0}%
\end{array}
\right)  .$ For this reason, without chiral symmetry, the wave-function of
edge state on left/right end no more completely localizes on A/B-sublattice.
The non-Hermitian similarity transformation on different sublattices, i.e.,
$\sigma_{x}\rightarrow(\sigma_{x})^{\beta}=\left(
\begin{array}
[c]{cc}%
{0} & e{^{\beta}}\\
e^{-\beta} & {0}%
\end{array}
\right)  $ cannot lead to the non-Hermitian similarity transformation on
different edge states, i.e., $\tau_{x}\rightarrow(\tau_{x})^{\beta}\neq \left(
\begin{array}
[c]{cc}%
{0} & e{^{\beta}}\\
e^{-\beta} & {0}%
\end{array}
\right)  .$ The defective edge states would become normal. Therefore, the
singular defective edge states with $\gamma=1$ are unstable against the
perturbation breaking the chiral symmetry.

\section{Anomalous BBC and defective edge states for 2D Chern insulator under
non-Hermitian similarity transformation}

\subsection{Hermitian deformed Qi-Wu-Zhang model}

Before discussing the defective edge states for 2D non-Hermitian Chern
insulator, we firstly consider a lattice model (deformed Qi-Wu-Zhang model) of
the 2D Hermitian Chern insulator. The Bloch Hamiltonian is given by
\begin{align}
H_{0}^{\mathrm{DQWZ}}(k_{x},k_{y})  &  =(v_{x}\sin k_{x})\sigma_{z}+(v_{y}\sin
k_{y})\sigma_{y}\nonumber \\
&  +(\mu-t_{x}\cos k_{x}-t_{y}\cos k_{y})\sigma_{x},
\end{align}
where $\sigma_{x,y,z}$ are Pauli matrices. In this paper, we set $t_{x}%
=t_{y}=v_{x}=v_{y}=1$. $\mu$ is the real mass parameter. The eigenvalues of
the Hamiltonian can be written as
\begin{equation}
E_{\pm}\left(  \vec{k}\right)  =\pm \sqrt{\left(  \mu-\cos k_{x}-\cos
k_{y}\right)  ^{2}+\left(  \sin k_{x}\right)  ^{2}+\left(  \sin k_{y}\right)
^{2}}.
\end{equation}

\subsubsection{Inversion symmetry and generalized chiral symmetry}

Now, there exists inversion symmetry for $H_{0}^{\mathrm{DQWZ}}(k_{x},k_{y})$
as
\begin{equation}
\mathcal{\hat{I}}H_{0}^{\mathrm{DQWZ}}(k_{x},k_{y})\mathcal{\hat{I}}%
^{-1}=H_{0}^{\mathrm{DQWZ}}(-k_{x},-k_{y}),
\end{equation}
where the inversion operator $\mathcal{\hat{I}}$ is $\sigma_{x}$. This
equation implies
\begin{equation}
E(\vec{k})=E(-\vec{k}).
\end{equation}
In addition, we have a generalized chiral symmetry at $k_{x}=0$, i.e.,
\begin{equation}
\sigma_{z}H^{\mathrm{DQWZ}}(k_{x}=0,k_{y})\sigma_{z}=-H^{\mathrm{DQWZ}}%
(k_{x}=0,k_{y}).
\end{equation}

\subsubsection{Total $Z_{2}$ topological invariant}

To characterize the 2D topological insulator, we introduce a new topological
invariant -- total $Z_{2}$ topological invariant,%
\begin{equation}
\eta=\eta_{\vec{k}=\mathbf{(}0,0)}\eta_{\vec{k}=\mathbf{(}0,\pi)}\eta_{\vec
{k}=\mathbf{(}\pi,0)}\eta_{\vec{k}=\mathbf{(}\pi,\pi)}.
\end{equation}
See below discussion.

There are four high symmetry points in momentum space, $\mathbf{(}0,0),$
$\mathbf{(}0,\pi)$, $\mathbf{(}\pi,0)$,$\  \mathbf{(}\pi,\pi)$, at which we
have
\begin{equation}
\mathcal{\hat{I}}\left \vert \psi(\vec{k}=(0/\pi,0/\pi))\right \rangle
=\left \vert \psi(\vec{k}=(0/\pi,0/\pi))\right \rangle .
\end{equation}
Thus, the Bloch Hamiltonian is divided into two parts
\begin{align}
H_{0}^{\mathrm{DQWZ}}(\vec{k})  &  =H_{0}^{\mathrm{DQWZ}}(\vec{k}\neq
(0/\pi,0/\pi))\nonumber \\
+H_{0}^{\mathrm{DQWZ}}(\vec{k}  &  =(0/\pi,0/\pi)).
\end{align}
For $\vec{k}=\mathbf{(}0,0)$, we have%
\begin{equation}
\frac{1}{2}\mathrm{Tr}[\mathcal{\hat{I}}\cdot H_{0}^{\mathrm{DQWZ}}(\vec
{k}=\mathbf{(}0,0))]=\mu-t_{x}-t_{y}=\mu-2;
\end{equation}
for $\vec{k}=\mathbf{(}0,\pi)$, we have%
\begin{equation}
\frac{1}{2}\mathrm{Tr}[\mathcal{\hat{I}}\cdot H_{0}^{\mathrm{DQWZ}}(\vec
{k}=\mathbf{(}0,\pi))]=\mu-t_{x}+t_{y}=\mu;
\end{equation}
for $\vec{k}=\mathbf{(}\pi,0)$, we have%
\begin{equation}
\frac{1}{2}\mathrm{Tr}[\mathcal{\hat{I}}\cdot H_{0}^{\mathrm{DQWZ}}(\vec
{k}=\mathbf{(}\pi,0))]=\mu+t_{x}-t_{y}=\mu;
\end{equation}
for $\vec{k}=\mathbf{(}\pi,\pi)$, we have%
\begin{equation}
\frac{1}{2}\mathrm{Tr}[\mathcal{\hat{I}}\cdot H_{0}^{\mathrm{DQWZ}}(\vec
{k}=\mathbf{(}\pi,\pi))]=\mu+t_{x}+t_{y}=\mu+2.
\end{equation}
\

To describe the topological structure of the deformed QWZ model $H_{0}%
^{\mathrm{DQWZ}}(k)$, we define four $Z_{2}$ topological invariants,
\begin{align}
\eta_{\vec{k}=\mathbf{(}0,0)}  &  =\frac{\mathrm{Tr}[\mathcal{\hat{I}}\cdot
H_{0}^{\mathrm{DQWZ}}[\vec{k}=\mathbf{(}0,0)\mathbf{]]}}{\left \vert
\mathrm{Tr}[\mathcal{\hat{I}}\cdot H_{0}^{\mathrm{DQWZ}}[\vec{k}%
=\mathbf{(}0,0)\mathbf{]]}\right \vert }=\frac{\mu-t_{x}-t_{y}}{\left \vert
\mu-t_{x}-t_{y}\right \vert },\text{ }\\
\eta_{\vec{k}=\mathbf{(}0,\pi)}  &  =\frac{\mathrm{Tr}[\mathcal{\hat{I}}\cdot
H_{0}^{\mathrm{DQWZ}}[\vec{k}=\mathbf{(}0,\pi)\mathbf{]]}}{\left \vert
\mathrm{Tr}[\mathcal{\hat{I}}\cdot H_{0}^{\mathrm{DQWZ}}[\vec{k}%
=\mathbf{(}0,\pi)\mathbf{]]}\right \vert }=\frac{\mu-t_{x}+t_{y}}{\left \vert
\mu-t_{x}+t_{y}\right \vert },\nonumber \\
\eta_{\vec{k}=\mathbf{(}\pi,0)}  &  =\frac{\mathrm{Tr}[\mathcal{\hat{I}}\cdot
H_{0}^{\mathrm{DQWZ}}[\vec{k}=\mathbf{(}\pi,0)]\mathbf{]}}{\left \vert
\mathrm{Tr}[\mathcal{\hat{I}}\cdot H_{0}^{\mathrm{DQWZ}}[\vec{k}=\mathbf{(}%
\pi,0)\mathbf{]]}\right \vert }=\frac{\mu+t_{x}-t_{y}}{\left \vert \mu
+t_{x}-t_{y}\right \vert },\nonumber \\
\eta_{\vec{k}=\mathbf{(}\pi,\pi)}  &  =\frac{\mathrm{Tr}[\mathcal{\hat{I}%
}\cdot H_{0}^{\mathrm{DQWZ}}[\vec{k}=\mathbf{(}\pi,\pi)\mathbf{]]}}{\left \vert
\mathrm{Tr}[\mathcal{\hat{I}}\cdot H_{0}^{\mathrm{DQWZ}}[\vec{k}=\mathbf{(}%
\pi,\pi)\mathbf{]]}\right \vert }=\frac{\mu+t_{x}+t_{y}}{\left \vert \mu
+t_{x}+t_{y}\right \vert }.\nonumber
\end{align}
We use the number $^{\prime}1^{\prime}$ to denote the case $\eta_{\vec
{k}=(0/\pi,0/\pi)}=-1$ and the number $^{\prime}0^{\prime}$ to denote the case
$\eta_{\vec{k}=(0/\pi,0/\pi)}=1.$ Hence, there are totally $16$ different
cases which represent $16$ different universal classes of topological states
denoted by $\left(  1111\right)  $,\ $\left(  1110\right)  $,\ $\left(
1101\right)  $,\ $\left(  1101\right)  $,\ $\left(  0111\right)  $,\ $\left(
1100\right)  $,\ $\left(  1001\right)  $,\ $\left(  0011\right)  $,\ $\left(
0110\right)  $,\ $\left(  0101\right)  $,\ $\left(  1010\right)  $,\ $\left(
1000\right)  $,\ $\left(  0100\right)  $,\ $\left(  0010\right)  $,\ $\left(
0001\right)  $,\ $\left(  0000\right)  $. The total $Z_{2}$ topological
invariant is defined as
\begin{align}
\eta &  =\eta_{\vec{k}=\mathbf{(}0,0)}\eta_{\vec{k}=\mathbf{(}0,\pi)}%
\eta_{\vec{k}=\mathbf{(}\pi,0)}\eta_{\vec{k}=\mathbf{(}\pi,\pi)}\nonumber \\
&  =\left \{
\begin{array}
[c]{l}%
+1,\text{ trivial phase}\\
-1,\text{ topological phase}%
\end{array}
\right.  .
\end{align}
Then, $\eta$ becomes the topological invariant to characterize the universal
properties of $16$ different topological orders, of which there are $8$
trivial phase : $\left(  1111\right)  $, $\left(  1100\right)  $,\ $\left(
1001\right)  $,\ $\left(  0011\right)  $,\ $\left(  0110\right)  $,\ $\left(
0101\right)  $,\ $\left(  1010\right)  $,\ $\left(  0000\right)  $,\textrm{
}and $8$ topological phase, $\left(  1110\right)  $,\ $\left(  1101\right)
$,\ $\left(  1101\right)  $,\ $\left(  0111\right)  $,\ $\left(  1000\right)
$,\ $\left(  0100\right)  $,\ $\left(  0010\right)  $,\ $\left(  0001\right)
$.

For above deformed Qi-Wu-Zhang model with $t_{x}=t_{y}=v_{x}=v_{y}=1,$ the
total $Z_{2}$ topological invariant is obtained as%
\begin{equation}
\eta=\eta_{\vec{k}=\mathbf{(}0,0)}\eta_{\vec{k}=\mathbf{(}0,\pi)}\eta_{\vec
{k}=\mathbf{(}\pi,0)}\eta_{\vec{k}=\mathbf{(}\pi,\pi)}=\frac{\mu^{2}%
-4}{\left \vert \mu^{2}-4\right \vert }.
\end{equation}
There are two phases, topological phase with $\eta=-1$ in the region of
$\mu^{2}<4$ and trivial phase with $\eta=1$ in the region of $\mu^{2}>4$. At
$\mu=\pm2,$ there exists a topological phase transition.

\subsubsection{Effective edge Hamiltonian}

In the topological phase ($\mu^{2}<4$), we firstly write down the effective
Hamiltonian of the edge states.

For open boundary condition along $y$-direction, the topological phase
exhibits edge modes localized on the boundaries. For the edge states with wave
vector $k_{x},$ we define the basis%
\begin{equation}
\left(
\begin{array}
[c]{c}%
\left \vert \mathrm{e}_{0,k_{x}}^{\text{L}}\right \rangle \\
\left \vert \mathrm{e}_{0,k_{x}}^{\text{R}}\right \rangle
\end{array}
\right)  .
\end{equation}
The effective Hamiltonian of edge states for Hermitian 2D Chern insulator is
\begin{equation}
\mathcal{\hat{H}}_{\mathrm{eff}}=\tau^{z}\varepsilon_{k_{x}}+\tau^{x}%
\Delta_{k_{x}},
\end{equation}
where $\varepsilon_{k}=\pm \sin k_{x}$ is the dispersion of the edge states of
semi-infinite system and $\Delta_{k_{x}}\sim(\mu-\cos k_{x})^{N_{y}}$ is the
tunneling strength where $N_{y}$ is the number of lattice sites along
$y$-direction. As a result, the energy levels are
\begin{equation}
\Delta E=\pm \sqrt{(\sin k_{x})^{2}+(\Delta_{k_{x}})^{2}}.
\end{equation}
In thermodynamic limit $N_{y}\rightarrow \infty$, $\Delta_{k_{x}}\rightarrow0,$
we have
\begin{equation}
\mathcal{\hat{H}}_{\mathrm{eff}}\rightarrow \tau^{z}\sin k_{x}.
\end{equation}

\subsection{Deformed Qi-Wu-Zhang model under non-Hermitian similarity
transformation}

Next, we consider the deformed QWZ model under non-Hermitian similarity
transformation $S=\left(
\begin{array}
[c]{cc}%
{1} & 0\\
0 & e^{\beta}%
\end{array}
\right)  $ where $\beta$ denotes non-Hermiticity strength.

Under the non-Hermitian similarity transformation, the Hamiltonian of deformed
QWZ model turns into
\begin{align}
H_{0}^{\mathrm{DQWZ}}  &  \rightarrow S^{-1}H_{0}^{\mathrm{DQWZ}%
}S=H^{\mathrm{DQWZ}}\nonumber \\
&  =(v_{x}\sin k_{x})\sigma_{z}+(v_{y}\sin k_{y})(\sigma_{y})^{\beta}\\
&  +(\mu-t_{x}\cos k_{x}-t_{y}\cos k_{y})(\sigma_{x})^{\beta}.\nonumber
\end{align}
The eigenvalues for the Hamiltonian $H^{\mathrm{DQWZ}}$ for the non-Hermitian
QWZ model are same to those for the Hamiltonian $H_{0}^{\mathrm{DQWZ}}$, i.e.,%
\begin{equation}
E_{\pm}\left(  k\right)  =\pm \sqrt{\left(  \mu-\cos k_{x}-\cos k_{y}\right)
^{2}+\left(  \sin k_{x}\right)  ^{2}+\left(  \sin k_{y}\right)  ^{2}}.
\end{equation}
Fig. 5 shows the energy spectra of deformed QWZ model under non-Hermitian
similarity transformation that are all real and unchanged with $\beta$.

\subsubsection{Inversion symmetry and generalized chiral symmetry}

For the non-Hermitian deformed QWZ model, there also exists the inversion
symmetry for $H^{\mathrm{DQWZ}}(k_{x},k_{y})$, i.e.,
\begin{equation}
\mathcal{\hat{I}}H^{\mathrm{DQWZ}}(k_{x},k_{y})\mathcal{\hat{I}}%
^{-1}=H^{\mathrm{DQWZ}}(-k_{x},-k_{y})
\end{equation}
where the inversion operator $\mathcal{\hat{I}}$ is $(\sigma_{x})^{\beta},$
and a generalized chiral symmetry at $k_{x}=0$, i.e.,
\begin{equation}
\sigma_{z}H^{\mathrm{DQWZ}}(k_{x}=0,k_{y})\sigma_{z}=-H^{\mathrm{DQWZ}}%
(k_{x}=0,k_{y}).
\end{equation}

\subsubsection{Total $Z_{2}$ topological invariant}

For non-Hermitian deformed QWZ model with inversion symmetry, we use the total
$Z_{2}$ topological invariant $\eta$ to characterize its topological
properties,%
\begin{align}
\eta &  =\eta_{\vec{k}=\mathbf{(}0,0)}\eta_{\vec{k}=\mathbf{(}0,\pi)}%
\eta_{\vec{k}=\mathbf{(}\pi,0)}\eta_{\vec{k}=\mathbf{(}\pi,\pi)}\nonumber \\
&  =\frac{\mathrm{Tr}[\mathcal{\hat{I}}\cdot H^{\mathrm{DQWZ}}[\vec
{k}=\mathbf{(}0,0)\mathbf{]]}}{\left \vert \mathrm{Tr}[\mathcal{\hat{I}}\cdot
H^{\mathrm{DQWZ}}[\vec{k}=\mathbf{(}0,0)\mathbf{]]}\right \vert }\nonumber \\
&  \cdot \frac{\mathrm{Tr}[\mathcal{\hat{I}}\cdot H^{\mathrm{DQWZ}}[\vec
{k}=\mathbf{(}0,\pi)]\mathbf{]}}{\left \vert \mathrm{Tr}[\mathcal{\hat{I}}\cdot
H^{\mathrm{DQWZ}}[\vec{k}=\mathbf{(}0,\pi \mathbf{)]]}\right \vert }\nonumber \\
&  \cdot \frac{\mathrm{Tr}[\mathcal{\hat{I}}\cdot H^{\mathrm{DQWZ}}[\vec
{k}=\mathbf{(}\pi,0)\mathbf{]]}}{\left \vert \mathrm{Tr}[\mathcal{\hat{I}}\cdot
H^{\mathrm{DQWZ}}[\vec{k}=\mathbf{(}\pi,0)\mathbf{]]}\right \vert }\nonumber \\
&  \cdot \frac{\mathrm{Tr}[\mathcal{\hat{I}}\cdot H^{\mathrm{DQWZ}}[\vec
{k}=\mathbf{(}\pi,\pi)]\mathbf{]}}{\left \vert \mathrm{Tr}[\mathcal{\hat{I}%
}\cdot H^{\mathrm{DQWZ}}[\vec{k}=\mathbf{(}\pi,\pi)\mathbf{]]}\right \vert }\\
&  =\frac{\mu^{2}-4}{\left \vert \mu^{2}-4\right \vert }.\nonumber
\end{align}
The global phase diagram doesn't change, i.e., topological phase with
$\eta=-1$ in the region of $\mu^{2}<4$ and trivial phase with $\eta=1$ in the
region of $\mu^{2}>4$. At $\mu=\pm2,$ there exists a topological phase transition.

\begin{figure}[ptb]
\includegraphics[clip,width=0.40\textwidth]{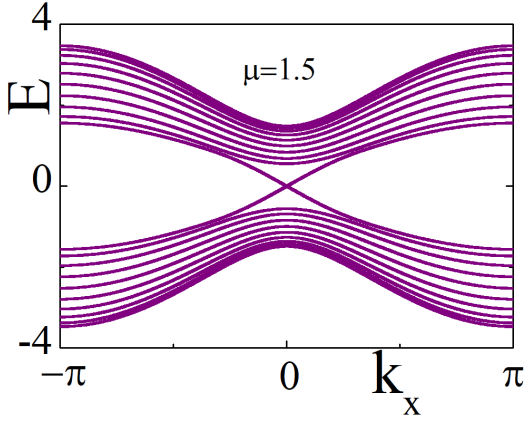}\caption{(Color online)
The energy spectra for 1D SSH model under non-Hermitian similarity
transformation. The energy spectra are independent on $\beta.$}%
\end{figure}

\subsubsection{Effective edge Hamiltonian for defective edge states}

In topological phase ($\mu^{2}<4$), the effective Hamiltonian of edge states
for the non-Hermitian 2D Chern insulator becomes
\begin{equation}
\mathcal{\breve{H}}_{\mathrm{eff}}=\tau^{z}\varepsilon_{k_{x}}+(\tau
^{x})^{\beta}\Delta_{k_{x}},
\end{equation}
where%
\begin{equation}
(\tau^{x})^{\beta}=\cosh(\beta)\tau^{x}+i\sinh(\beta)\tau^{y}=\left(
\begin{array}
[c]{cc}%
{0} & {\mathrm{e}^{\beta}}\\
{\mathrm{e}^{-\beta}} & {0}%
\end{array}
\right)  .
\end{equation}
This effective edge Hamiltonian is also non-Hermitian, i.e.,
\begin{equation}
\mathcal{\breve{H}}_{\mathrm{eff}}\neq \mathcal{\breve{H}}_{\mathrm{eff}%
}^{\dagger}.
\end{equation}
The energy levels for the edge states are
\begin{equation}
\Delta E=\pm \sqrt{(\sin k_{x})^{2}+(\Delta_{k_{x}})^{2}}%
\end{equation}
that are same to the Hermitian case with $\beta=0$. In Fig. 5, the spectra of
edge states are given.

However, the basis for the edge states changes under non-Hermitian similarity
transformation, i.e., for the edge states with wave vector $k_{x}$
\begin{align}
\left(
\begin{array}
[c]{c}%
\left \vert \mathrm{e}_{0,k_{x}}^{\text{L}}\right \rangle \\
\left \vert \mathrm{e}_{0,k_{x}}^{\text{R}}\right \rangle
\end{array}
\right)   &  \rightarrow \left(
\begin{array}
[c]{c}%
\left \vert \mathrm{\bar{e}}_{0,k_{x}}^{\text{L}}\right \rangle \\
\left \vert \mathrm{\bar{e}}_{0,k_{x}}^{\text{R}}\right \rangle
\end{array}
\right)  =S\left(
\begin{array}
[c]{c}%
\left \vert \mathrm{e}_{0,k_{x}}^{\text{L}}\right \rangle \\
\left \vert \mathrm{e}_{0,k_{x}}^{\text{R}}\right \rangle
\end{array}
\right) \\
&  =\left(
\begin{array}
[c]{c}%
\left \vert \mathrm{e}_{0,k_{x}}^{\text{L}}\right \rangle \\
{\mathrm{e}^{-\beta}}\left \vert \mathrm{e}_{0,k_{x}}^{\text{R}}\right \rangle
\end{array}
\right)  ,\nonumber
\end{align}
where $\left(
\begin{array}
[c]{c}%
\left \vert \mathrm{\bar{e}}_{0,k_{x}}^{\text{L}}\right \rangle \\
\left \vert \mathrm{\bar{e}}_{0,k_{x}}^{\text{R}}\right \rangle
\end{array}
\right)  $ is the basis for the non-Hermitian case with $\beta \neq0$. To
characterize the non-Hermitian properties from similarity transformation, we
also define the states overlap $\gamma_{k_{x}}$ between the two edge states
with wave vector $k_{x}$ to be
\begin{equation}
\gamma_{k_{x}}=\left \langle \psi_{k_{x},2}|\psi_{k_{x},1}\right \rangle
=\tanh \beta.
\end{equation}
For the case of $\beta \rightarrow0$, we have
\begin{equation}
\left(
\begin{array}
[c]{c}%
\left \vert \mathrm{\bar{e}}_{0,k_{x}}^{\text{L}}\right \rangle \\
\left \vert \mathrm{\bar{e}}_{0,k_{x}}^{\text{R}}\right \rangle
\end{array}
\right)  \rightarrow \left(
\begin{array}
[c]{c}%
\left \vert \mathrm{e}_{0,k_{x}}^{\text{L}}\right \rangle \\
\left \vert \mathrm{e}_{0,k_{x}}^{\text{R}}\right \rangle
\end{array}
\right)  .
\end{equation}
Now, we have $\gamma_{k_{x}}\rightarrow0;$ On the other hand, for the case of
$\beta \rightarrow \infty$, we have
\begin{equation}
\left(
\begin{array}
[c]{c}%
\left \vert \mathrm{\bar{e}}_{0,k_{x}}^{\text{L}}\right \rangle \\
\left \vert \mathrm{\bar{e}}_{0,k_{x}}^{\text{R}}\right \rangle
\end{array}
\right)  \rightarrow \left(
\begin{array}
[c]{c}%
\left \vert \mathrm{e}_{0,k_{x}}^{\text{L}}\right \rangle \\
0
\end{array}
\right)  .
\end{equation}
Now, we have $\gamma_{k_{x}}\rightarrow1.$ All edge states with different wave
vectors become defective: only edge states at left or right boundary exists.

\begin{figure}[ptb]
\includegraphics[clip,width=0.48\textwidth]{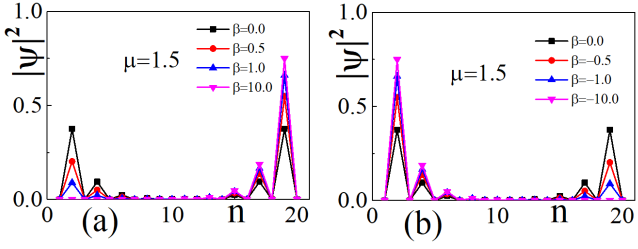}\caption{(Color online)
The wave-functions via $\beta$ for edge states of 2D deformed QWZ model under
non-Hermitian similarity transformation: $\beta>0$ for (a) and $\beta<0$\ for
(b). For this case, we have $t_{x}=t_{y}=v_{x}=v_{y}=1.0$, at $\mu=1.5$.}%
\label{Fig.4}%
\end{figure}

Fig. 6(a), 6(b) show the edge states along $y$-direction of deformed QZW model
under non-Hermitian similarity transformation. One can see that the
non-Hermitian similarity transformation will polarize the states spin in
degrees of freedom. In the strong non-Hermitian limit $\beta \rightarrow
\pm \infty,$ the edge states are polarized in spin degrees of freedom that
corresponds to a defective edge states on left/right end. Fig. 7 shows the
states overlap $\gamma_{k_{x}}$ between the edge states $\psi_{k_{x},1}$ and
$\psi_{k_{x},2}$ with the same wave vector $k_{x}=0.3.$ From Fig. 7, one can
see that the theoretical prediction is also consistent to the numerical
results. In the strong non-Hermitian limit $\beta \rightarrow \pm \infty,$
$\gamma_{k_{x}}=\left \langle \psi_{k_{x},2}|\psi_{k_{x},1}\right \rangle
$\ turns to $1$. This indicates the (singular) defective edge states.
\begin{figure}[ptb]
\includegraphics[clip,width=0.40\textwidth]{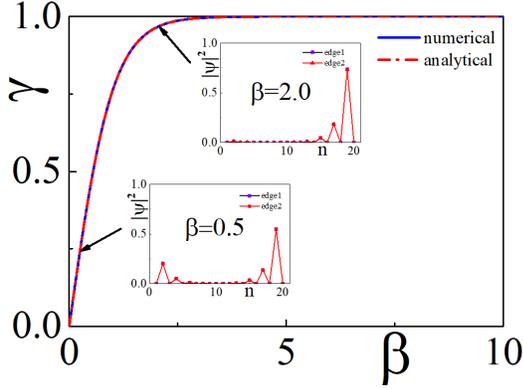}\caption{(Color online)
The state overlap $\gamma$ between the two edge states with same wave vector
$k_{x}=0.3$ for deformed QZW model under non-Hermitian similarity
transformation $.$ }%
\end{figure}

For 2D non-Hermitian topological insulators, the intrinsic defective edge
states have not been discovered yet. This is an example (in particular, in the
strong non-Hermitian limit $\beta \rightarrow \pm \infty$), of which there only
exist edge states on left or right end. In addition, the defective edge states
are protected by a generalized chiral symmetry. For the edge states along
$y$-direction without a generalized chiral symmetry for the edge states at
$k_{y}=0,$ i.e.,%
\begin{equation}
\sigma_{z}H^{\mathrm{DQWZ}}(k_{x},k_{y}=0)\sigma_{z}\neq-H^{\mathrm{DQWZ}%
}(k_{x},k_{y}=0),
\end{equation}
there doesn't exist defective edge states any more.

\subsection{Usual bulk-boundary correspondence for traditional non-Hermitian
Qi-Wu-Zhang model under ST}

In addition, we consider the traditional non-Hermitian Qi-Wu-Zhang model under
similarity transformation. The Bloch Hamiltonian is
\begin{align}
H^{\mathrm{TQWZ}}  &  =(v_{x}\sin k_{x})(\sigma_{x})^{\beta}+(v_{y}\sin
k_{y})(\sigma_{y})^{\beta}\\
&  +(\mu-t_{x}\cos k_{x}-t_{y}\cos k_{y})\sigma_{z}\nonumber
\end{align}
The energy spectra for the bulk states of this Hamiltonian can be written as
\begin{equation}
E_{\pm}\left(  k\right)  =\pm \sqrt{\left(  \mu-\cos k_{x}-\cos k_{y}\right)
^{2}+\left(  \sin k_{x}\right)  ^{2}+\left(  \sin k_{y}\right)  ^{2}}.
\end{equation}

For the non-Hermitian traditional QWZ model, there also exists the inversion
symmetry for $H^{\mathrm{DQWZ}}(k_{x},k_{y})$, i.e.,
\begin{equation}
\mathcal{\hat{I}}H^{\mathrm{DQWZ}}(k_{x},k_{y})\mathcal{\hat{I}}%
^{-1}=H^{\mathrm{DQWZ}}(-k_{x},-k_{y})
\end{equation}
where the inversion operator $\mathcal{\hat{I}}$ is $\sigma_{z}.$ However,
there doesn't exist a generalized chiral symmetry at $\vec{k}=0$, i.e.,
\begin{align}
\sigma_{z}H^{\mathrm{TQWZ}}(k_{x}  &  =0,k_{y})\sigma_{z}\neq-H^{\mathrm{TQWZ}%
}(k_{x}=0,k_{y}),\nonumber \\
\sigma_{z}H^{\mathrm{TQWZ}}(k_{x},k_{y}  &  =0)\sigma_{z}\neq-H^{\mathrm{TQWZ}%
}(k_{x},k_{y}=0).
\end{align}

For traditional Qi-Wu-Zhang model under non-Hermitian similarity
transformation, the total $Z_{2}$ topological invariant can be also used to
characterize its topological properties,%
\begin{align}
\eta &  =\eta_{\vec{k}=\mathbf{(}0,0)}\eta_{\vec{k}=\mathbf{(}0,\pi)}%
\eta_{\vec{k}=\mathbf{(}\pi,0)}\eta_{\vec{k}=\mathbf{(}\pi,\pi)}\nonumber \\
&  =\frac{\mathrm{Tr}[\mathcal{\hat{I}}\cdot H^{\mathrm{TQWZ}}[\vec
{k}=\mathbf{(}0,0)]\mathbf{]}}{\left \vert \mathrm{Tr}[\mathcal{\hat{I}}\cdot
H_{0}^{\mathrm{TQWZ}}[\vec{k}=\mathbf{(}0,0)\mathbf{]]}\right \vert
}\nonumber \\
&  \cdot \frac{\mathrm{Tr}[\mathcal{\hat{I}}\cdot H^{\mathrm{TQWZ}}[\vec
{k}=\mathbf{(}0,\pi)]\mathbf{]}}{\left \vert \mathrm{Tr}[\mathcal{\hat{I}}\cdot
H^{\mathrm{TQWZ}}[\vec{k}=\mathbf{(}0,\pi)\mathbf{]]}\right \vert }\nonumber \\
&  \cdot \frac{\mathrm{Tr}[\mathcal{\hat{I}}\cdot H^{\mathrm{TQWZ}}[\vec
{k}=\mathbf{(}\pi,0)\mathbf{]]}}{\left \vert \mathrm{Tr}[\mathcal{\hat{I}}\cdot
H^{\mathrm{TQWZ}}[\vec{k}=\mathbf{(}\pi,0)\mathbf{]]}\right \vert }\nonumber \\
&  \cdot \frac{\mathrm{Tr}[\mathcal{\hat{I}}\cdot H^{\mathrm{TQWZ}}[\vec
{k}=\mathbf{(}\pi,\pi)]\mathbf{]}}{\left \vert \mathrm{Tr}[\mathcal{\hat{I}%
}\cdot H^{\mathrm{TQWZ}}[\vec{k}=\mathbf{(}\pi,\pi)\mathbf{]]}\right \vert }\\
&  =\frac{\mu^{2}-4}{\left \vert \mu^{2}-4\right \vert }.\nonumber
\end{align}
There are also two phases, topological phase with $\eta=-1$ in the region of
$\mu^{2}<4$ and trivial phase with $\eta=1$ in the region of $\mu^{2}>4$. At
$\mu=\pm2,$ there exists a topological phase transition.

In topological phase, the effective Hamiltonian of edge states becomes
\begin{equation}
\mathcal{\breve{H}}_{\mathrm{eff}}=\tau^{z}\varepsilon_{k_{x}}+\tau^{x}%
\Delta_{k_{x}}.
\end{equation}
This effective edge Hamiltonian is Hermitian, i.e.,
\begin{equation}
\mathcal{\breve{H}}_{\mathrm{eff}}=\mathcal{\breve{H}}_{\mathrm{eff}}%
^{\dagger}.
\end{equation}
The energy levels are
\begin{equation}
\Delta E=\pm \sqrt{(\sin k_{x})^{2}+(\Delta_{k_{x}})^{2}}%
\end{equation}

Except for the non-Hermitian spin polarization effect, the basis for the edge
states doesn't change under non-Hermitian similarity transformation, i.e., for
the edge states with wave vector $k_{x}$
\begin{equation}
\left(
\begin{array}
[c]{c}%
\left \vert \mathrm{e}_{0,k_{x}}^{\text{L}}\right \rangle \\
\left \vert \mathrm{e}_{0,k_{x}}^{\text{R}}\right \rangle
\end{array}
\right)  \rightarrow \left(
\begin{array}
[c]{c}%
\left \vert \mathrm{\bar{e}}_{0,k_{x}}^{\text{L}}\right \rangle \\
\left \vert \mathrm{\bar{e}}_{0,k_{x}}^{\text{R}}\right \rangle
\end{array}
\right)  .
\end{equation}
Now, the state overlap $\gamma_{k_{x}}$ between the two edge states with wave
vector $k_{x}$ is trivial,
\begin{equation}
\gamma_{k_{x}}=\left \langle \psi_{k_{x},2}|\psi_{k_{x},1}\right \rangle
\equiv0.
\end{equation}

Then we plot the wave-function for edge states under open boundary condition
in Fig. 8(a). We find that the wave-function will never\ localize on one edge
and there is no defective edge state. Instead, the edge states have
distribution on both edges with non-Hermitian spin polarization effect. Fig.
8(b) shows the overlap $\gamma=$ $\left \langle \psi_{k_{x},2}|\psi_{k_{x}%
,1}\right \rangle $ between the two edge states with wave vector $k_{x}=0.3.$
From Fig. 8(b), one can see that the theoretical prediction is also consistent
to the numerical results, i.e., $\left \langle \psi_{k_{x},2}|\psi_{k_{x}%
,1}\right \rangle \equiv0$. This indicates the usual edge states.
\begin{figure}[ptb]
\includegraphics[clip,width=0.48\textwidth]{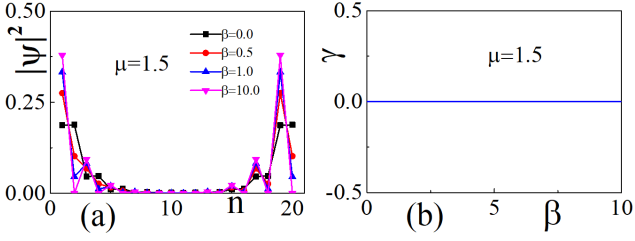}\caption{(Color online)
(a) The wave-functions via $\beta$ for edge states of 2D traditional QWZ model
under non-Hermitian similarity transformation. For this case, we have
$t_{x}=t_{y}=v_{x}=v_{y}=1.0$, at $\mu=1.5$; (b) The state overlap $\gamma$
between the two edge states.}%
\end{figure}

\section{Anomalous BBC and defective edge states for topological insulator on
the $d$-D cubic lattice under non-Hermitian similarity transformation}

In above sections, we have discussed the anomalous BBC\ and defective edge
states for 1D SSH model and 2D (deformed) QWZ under non-Hermitian similarity
transformation. In this section, we will generalize the results to topological
insulators in higher dimensions under non-Hermitian similarity transformation.

\subsection{Non-Hermitian topological insulators on the $d$-D cubic lattice}

Assuming periodic boundary conditions along all directions, we consider the
following Hamiltonian for non-Hermitian topological insulators on the $d$-D
cubic lattice
\begin{align}
H_{\mathrm{cubic}}(\vec{k})  &  =\sum_{i\neq j}^{d}\sin k_{i}\Gamma_{i}+\sin
k_{j}(\Gamma_{j})^{\beta}\nonumber \\
&  +(\mu-\sum_{i=1}^{d}\cos k_{i})(\Gamma_{d+1})^{\beta},
\end{align}
where $\Gamma_{\mu}$ denotes the gamma matrices that satisfy $\{ \Gamma_{\mu
},\Gamma_{\nu}\}=2\delta_{\mu \nu}$ and $\mu$ is the real mass parameter. Here,
we have
\begin{align}
\Gamma_{j}  &  \rightarrow(\Gamma_{j})^{\beta}=\cosh(\beta)\Gamma_{j}%
+i\sinh(\beta)\Gamma_{d+1},\\
\Gamma_{d+1}  &  \rightarrow(\Gamma_{d+1})^{\beta}=\cosh(\beta)\Gamma
_{d+1}-i\sinh(\beta)\Gamma_{j}.\nonumber
\end{align}
The energy spectra for the bulk states of this Hamiltonian can be written as
\begin{equation}
E_{\pm}\left(  \vec{k}\right)  =\pm \sqrt{\left(  \mu-\sum_{i=1}^{d}\cos
k_{i}\right)  ^{2}+\sum_{i}^{d}\sin^{2}k_{i}}.
\end{equation}

\subsection{Inversion symmetry and generalized chiral symmetry}

There exists inversion symmetry for $H_{\mathrm{cubic}}(\vec{k})$ as
\begin{equation}
\mathcal{\hat{I}}H_{\mathrm{cubic}}(\vec{k})\mathcal{\hat{I}}^{-1}%
=H_{\mathrm{cubic}}(-\vec{k}),
\end{equation}
where the inversion operator $\mathcal{\hat{I}}$ is $(\Gamma_{d+1})^{\beta}$.
In addition, we have a generalized chiral symmetry for the edge states at
$k_{i\neq j}=0$, i.e.,
\begin{equation}
\Gamma_{j}H_{\mathrm{cubic}}(\vec{k})(k_{i\neq j}=0,k_{j})\Gamma
_{j}=-H_{\mathrm{cubic}}(\vec{k})\text{ }(k_{i\neq j}=0,k_{j}).
\end{equation}

\subsection{Total $Z_{2}$ topological invariant}

There are $2^{d}$ high symmetry points in momentum space, $\mathbf{(}%
0,...,0),$ $\mathbf{(}0,...,\pi)$, $\mathbf{(}0,...\pi,0)$,$\ $..., at which
we have
\begin{equation}
\mathcal{\hat{I}}\left \vert \psi(\vec{k}=(0/\pi,...,0/\pi))\right \rangle
=\left \vert \psi(\vec{k}=(0/\pi,...,0/\pi))\right \rangle .
\end{equation}
To characterize the non-Hermitian topological insulators on the $d$-D cubic
lattice, we introduce a new topological invariant -- total $Z_{2}$ topological
invariant,
\begin{equation}
\eta=%
{\displaystyle \prod \limits_{\vec{k}=\mathbf{0}}}
\eta_{\vec{k}=\mathbf{0}},
\end{equation}
where $\mathbf{0}$ means the $2^{d}$ high symmetry points in momentum space
$\mathbf{(}0,...,0),$ $\mathbf{(}0,...,\pi)$, $\mathbf{(}0,...\pi,0)$,$\ $...

The Bloch Hamiltonian is divided into two parts
\begin{align}
H_{\mathrm{cubic}}(\vec{k})  &  =H_{\mathrm{cubic}}(\vec{k}\neq(0/\pi
,...,0/\pi))\nonumber \\
+H_{\mathrm{cubic}}(\vec{k}  &  =(0/\pi,...,0/\pi))
\end{align}
\ To describe this topological structure of $H_{\mathrm{cubic}}(\vec{k})$, we
define $2^{d}$ $Z_{2}$ topological invariants,
\begin{equation}
\eta_{\vec{k}=\mathbf{0}}=\frac{\mathrm{Tr}[\mathcal{\hat{I}}\cdot
H_{\mathrm{cubic}}[\vec{k}=\mathbf{0]]}}{\left \vert \mathrm{Tr}[\mathcal{\hat
{I}}\cdot H_{\mathrm{cubic}}[\vec{k}=\mathbf{0]]}\right \vert }.\text{ }%
\end{equation}
For example, $\vec{k}=\mathbf{(}0,...,0)$, we have%
\begin{equation}
\mathrm{Tr}[\mathcal{\hat{I}}\cdot H_{\mathrm{cubic}}(\vec{k}=\mathbf{(}%
0,...,0))]=2(\mu-d).
\end{equation}
Now, we use the number $^{\prime}1^{\prime}$ to denote the case $\eta_{\vec
{k}=\mathbf{0}}=-1$ and the number $^{\prime}0^{\prime}$ to denote the case
$\eta_{\vec{k}=\mathbf{0}}=1.$ Hence, there are totally $2^{d}$ different
cases which represent $2^{2^{d}}$ different universal classes of topological states.

The total $Z_{2}$ topological invariant is defined as
\begin{align}
\eta &  =%
{\displaystyle \prod \limits_{\vec{k}\mathbf{=0}}}
\eta_{\vec{k}=\mathbf{0}}=%
{\displaystyle \prod \limits_{\vec{k}=\mathbf{0}}}
\frac{\mathrm{Tr}[\mathcal{\hat{I}}\cdot H_{\mathrm{cubic}}[\vec
{k}=\mathbf{0]]}}{\left \vert \mathrm{Tr}[\mathcal{\hat{I}}\cdot
H_{\mathrm{cubic}}[\vec{k}=\mathbf{0]]}\right \vert }\nonumber \\
&  =\left \{
\begin{array}
[c]{l}%
+1,\text{ trivial phase}\\
-1,\text{ topological phase}%
\end{array}
\right.  .
\end{align}
Then, $\eta$ becomes topological invariant to characterize the universal
properties of different topological orders, of which there are $2^{2^{d}-1}$
trivial phases for the case of $\mu^{2}>d^{2}$,\textrm{ }and $2^{2^{d}-1}$
topological phases, $\mu^{2}<d^{2}$. At $\mu=\pm d,$ there exists a
topological phase transition.

\subsection{Effective edge Hamiltonian}

In topological phase ($\left \vert \mu \right \vert <\left \vert d\right \vert $),
there exists gapless edge states. Along $x_{j}$-direction, the effective edge
Hamiltonian becomes
\begin{equation}
\mathcal{\breve{H}}_{\mathrm{eff}}=\Gamma_{i}\otimes \tau^{z}(\sum_{i\neq
j}\sin k_{i})+\Gamma_{i}\otimes(\tau^{x})^{\beta}\Delta_{k_{i}}%
\end{equation}
where $\Delta_{k_{i}}\sim(\mu-\sum_{i\neq j}\cos k_{i})^{N_{y}}$, and
\begin{equation}
(\tau^{x})^{\beta}=\cosh(\beta)\tau^{x}+i\sinh(\beta)\tau^{y}=\left(
\begin{array}
[c]{cc}%
{0} & {\mathrm{e}^{\beta}}\\
{\mathrm{e}^{-\beta}} & {0}%
\end{array}
\right)  .
\end{equation}
This effective edge Hamiltonian is non-Hermitian, i.e.,
\begin{equation}
\mathcal{\breve{H}}_{\mathrm{eff}}\neq \mathcal{\breve{H}}_{\mathrm{eff}%
}^{\dagger}.
\end{equation}
In thermodynamic limit $N_{j}\rightarrow \infty$, $\Delta_{k_{i}}\rightarrow0,$
we have
\begin{equation}
\mathcal{\check{H}}_{\mathrm{eff}}\rightarrow \Gamma_{i}\otimes \tau^{z}%
(\sum_{i\neq j}\sin k_{i}).
\end{equation}
As a result, the energy levels for edge states become
\begin{equation}
\Delta E=\pm \sqrt{(\sum_{i\neq j}\sin k_{i})+(\Delta_{k_{i}})^{2}}.
\end{equation}
This indicates that the energy spectra for edge states are independent on
$\beta$.

\subsection{Defective edge states}

The basis for the edge states with wave vector $k_{i}$ will be changed under
the similarity transformation $S^{-1}$, i.e.,
\begin{align}
\left(
\begin{array}
[c]{c}%
\left \vert \mathrm{e}_{0,k_{i}}^{\text{L}}\right \rangle \\
\left \vert \mathrm{e}_{0,k_{i}}^{\text{R}}\right \rangle
\end{array}
\right)   &  \rightarrow \left(
\begin{array}
[c]{c}%
\left \vert \mathrm{\bar{e}}_{0,k_{i}}^{\text{L}}\right \rangle \\
\left \vert \mathrm{\bar{e}}_{0,k_{i}}^{\text{R}}\right \rangle
\end{array}
\right)  =S^{-1}\left(
\begin{array}
[c]{c}%
\left \vert \mathrm{e}_{0,k_{i}}^{\text{L}}\right \rangle \\
\left \vert \mathrm{e}_{0,k_{i}}^{\text{R}}\right \rangle
\end{array}
\right) \\
&  =\left(
\begin{array}
[c]{c}%
\left \vert \mathrm{e}_{0,k_{i}}^{\text{L}}\right \rangle \\
{\mathrm{e}^{-\beta}}\left \vert \mathrm{e}_{0,k_{i}}^{\text{R}}\right \rangle
\end{array}
\right) \nonumber
\end{align}
where for the edge states with wave vector $k_{i},$ $\left(
\begin{array}
[c]{c}%
\left \vert \mathrm{e}_{0,k_{i}}^{\text{L}}\right \rangle \\
\left \vert \mathrm{e}_{0,k_{i}}^{\text{R}}\right \rangle
\end{array}
\right)  $ is the basis for the Hermitian case with $\beta=0$ and $\left(
\begin{array}
[c]{c}%
\left \vert \mathrm{\bar{e}}_{0,k_{i}}^{\text{L}}\right \rangle \\
\left \vert \mathrm{\bar{e}}_{0,k_{i}}^{\text{R}}\right \rangle
\end{array}
\right)  $ is the basis for the non-Hermitian case with $\beta \neq0$.

To characterize the non-Hermitian properties from similarity transformation,
we define the state overlap $\gamma_{k_{i}}$ between the two edge states with
wave vector $k_{i}$ to be
\begin{equation}
\gamma_{k_{i}}=\left \langle \psi_{k_{i},2}|\psi_{k_{i},1}\right \rangle
=\tanh \beta.
\end{equation}
For the case of $\beta \rightarrow0$, we have
\[
\left(
\begin{array}
[c]{c}%
\left \vert \mathrm{\bar{e}}_{0,k_{i}}^{\text{L}}\right \rangle \\
\left \vert \mathrm{\bar{e}}_{0,k_{i}}^{\text{R}}\right \rangle
\end{array}
\right)  \rightarrow \left(
\begin{array}
[c]{c}%
\left \vert \mathrm{e}_{0,k_{i}}^{\text{L}}\right \rangle \\
\left \vert \mathrm{e}_{0,k_{i}}^{\text{R}}\right \rangle
\end{array}
\right)  .
\]
Now, we have $\gamma_{k_{i}}\rightarrow0;$ On the other hand, for the case of
$\beta \rightarrow \infty$, we have
\begin{equation}
\left(
\begin{array}
[c]{c}%
\left \vert \mathrm{\bar{e}}_{0,k_{i}}^{\text{L}}\right \rangle \\
\left \vert \mathrm{\bar{e}}_{0,k_{i}}^{\text{R}}\right \rangle
\end{array}
\right)  \rightarrow \left(
\begin{array}
[c]{c}%
\left \vert \mathrm{e}_{0,k_{i}}^{\text{L}}\right \rangle \\
0
\end{array}
\right)  .
\end{equation}
Now, we have $\gamma_{k_{i}}\rightarrow1.$ All edge states become defective:
only edge states at left or right boundary.

However, for the edge states along other directions, without generalized
chiral symmetry for the edge states at $k_{i}=0,$ the edge states become normal.

\section{Conclusion}

At the end, we give a brief conclusion. We have exhaustively analyzed a new
class of non-Hermitian topological systems: topological insulators under
non-Hermitian similarity transformation by using 1D SSH model, 2D (Deformed)
QWZ model as examples. With the help of a new type of symmetry-protected
topological invariant -- total $Z_{2}$ topological invariant $\eta=%
{\displaystyle \prod \limits_{\vec{k}=\mathbf{0}}}
\eta_{\vec{k}=\mathbf{0}}$ ($\mathbf{0}$ denotes $2^{d}$ high symmetry points
in momentum space), the topological phases and trivial phases are classified.
In topological phases without non-Hermitian skin effect, there exist defective
edge states that are protected by (general) chiral symmetry. The effective
edge Hamiltonian $\mathcal{\check{H}}_{\mathrm{eff}}$ are obtained to describe
the underlying physics of the defective edge states. The defectiveness of the
edge states can be verified by calculating the state overlap $\gamma=$
$\left \langle \psi_{2}|\psi_{1}\right \rangle $ between two edge states.

In addition, we point out that the symmetry-protected topological invariant --
total $Z_{2}$ topological invariant $\eta$ and the quantitative theory for the
defective edge states can be generalized to various types of non-Hermitian
topological systems, such as non-Hermitian topological superconductors and
non-Hermitian topological semi-metals. These issues will be studied in the future.

\acknowledgments This work is supported by NSFC Grant No. 11674026, 11974053.

\end{document}